\documentclass[pre, amssymb, amsmath, nofootinbib, 10pt, twocolumn]{revtex4-1}

	\usepackage{graphicx}
	\usepackage{color}
	\usepackage{verbatim}
	\usepackage{subfigure}
	\usepackage{amsmath}
	\usepackage{amssymb}

\begin{document}
	
	\title{Hybrid Models of Molecular Machines and the No-Pumping Theorem}
	\author{Dibyendu Mandal$^1$ and Christopher Jarzynski$^2$}
	\affiliation{
	$^1$Department of Physics, University of Maryland, College Park, MD 20742, U.S.A.\\ 
	$^2$Department of Chemistry and Biochemistry, and Institute of Physical Science and Technology, University of Maryland, College Park, MD 20742, U.S.A.}
	
	\begin{abstract}
	
Synthetic nanoscale complexes capable of mechanical movement are often studied theoretically using discrete-state models that involve instantaneous transitions between metastable states. 
A number of general results have been derived within this framework, including a ``no-pumping theorem" that restricts the possibility of generating directed motion by the periodic variation of external parameters. 
Motivated by recent experiments using time-resolved vibrational spectroscopy [Panman et al., Science 328, 1255 (2010)], we introduce a more detailed and realistic class of models in which transitions between metastable states occur by finite-time, diffusive processes rather than sudden jumps. 
We show that the no-pumping theorem remains valid within this framework.
 	
	\end{abstract}
	
	\maketitle

\section{Introduction}

Synthetic molecular complexes capable of mechanical movement, often simply called {\it molecular machines}, have caught the attention of researchers in the past decade for their potential applications in nanomechanical devices \cite{Bath2007, Michl2009}.
Numerous designs have been proposed and experimentally realized, including unidirectional rotation of molecular components \cite{Kelly1999, Koumura1999}, molecular translation on tracks \cite{Delius2009} and surfaces \cite{Lund2010}, a programmable DNA nanoscale assembly line \cite{Gu2010} and a single molecule electric motor \cite{Tierney2011}.
The operating principles of these molecular machines differ from those of their macroscopic counterparts: inertial effects are negligible, whereas thermal fluctuations play a dominant role, making their description inherently stochastic.

A number of recent theoretical studies have focused on the response of molecular machines to the time-periodic variation of external parameters, a mode of operation often called {\it stochastic pumping}.
In this setting, the currents induced in the molecular machine have been shown to have geometric contributions \cite{Parrondo1998, Astumian2003, Sinitsyn2007, Ohkubo2008, Sinitsyn2009} analogous to the Berry and Aharonov-Anandan phases in quantum mechanics \cite{Berry1984,Aharonov1987}.
Moreover, in Ref.\cite{Rahav2008} an exact expression for the average pumped current was derived, and was used to establish a {\it no-pumping theorem} (NPT), a restrictive condition for generating directed motion in stochastic pumps.
These results were further studied and extended in Refs.~\cite{Horowitz2009, Chernyak2008, Maes2010, Ren2011, Mandal2011, Sinitsyn2011,  Chernyak2012JCPa, Chernyak2012JCPb}.  
Most of this theoretical work has involved discrete-state models in which the molecular machine makes instantaneous, thermally activated transitions among a set of metastable states.
Such models are conveniently visualized using a network representation (see e.g. Fig.~\ref{fig:currentsgen}), where the nodes of a graph depict the metastable states and the edges indicate the allowed sudden transitions \cite{Schnakenberg1976, Hill1989}.

In reality, a transition between two metastable states of a molecular machine involves mechanical motion, and therefore can not be instantaneous.
Recent experiments \cite{Panman2010,Panman2012} using time-resolved vibrational spectroscopy to study the movement of a molecular machine between two docking stations provide evidence that this motion is described more accurately as a rapid, one-dimensional random walk than as an instantaneous jump.
This random walk description can capture the effects of physical separation between the docking stations.
Motivated by this observation, in this paper we introduce a model of molecular machines in which the system makes diffusive (rather than sudden) transitions.
The metastable states continue to be depicted by the nodes of a graph, but during a transition from one state to another the machine evolves diffusively along the connecting edge.
Thus the machine now lives in a hybrid state space, consisting of both discrete and continuous components, represented by the nodes and edges of the graph.

Because our hybrid model aims to provide a more detailed microscopic description of the dynamics of molecular machines, it is natural to wonder whether the theoretical results mentioned earlier, which were derived under the assumption of instantaneous state-to-state transitions, remain valid within the framework we now propose.
In this paper we will focus specifically on the no-pumping theorem (NPT) of Ref.~\cite{Rahav2008}, which was observed in actual experiments \cite{Leigh2003}, in fact prior to its general theoretical formulation.
We will show that the NPT remains exactly valid within our hybrid model.

The no-pumping theorem was experimentally observed in Ref.~\cite{Leigh2003} for a [2]catenane complex, a rotatory molecular machine.
The [2]catenane also provides a minimal setup for illustrating key elements of our hybrid framework and its analysis. 
The first part of the text is therefore devoted to the discussion of this particular system.
We then generalize our framework by extending it to more complex molecular machines.
The organization of the paper is as follows: Sec.~\ref{sec:model} introduces the hybrid model for the [2]catenane; Sec.~\ref{sec:DB} imposes detailed balance; Sec.~\ref{sec:NPT} provides the proof of the no-pumping theorem; and Sec.~\ref{sec:generalization} generalizes these discussions to arbitrary systems. 

\section{Hybrid Model of a [2]catenane}
	\label{sec:model}
	
	\begin{figure}[tbp]
	\centering
	\includegraphics[trim = 0in 0in 0in 0in, scale= 0.65]{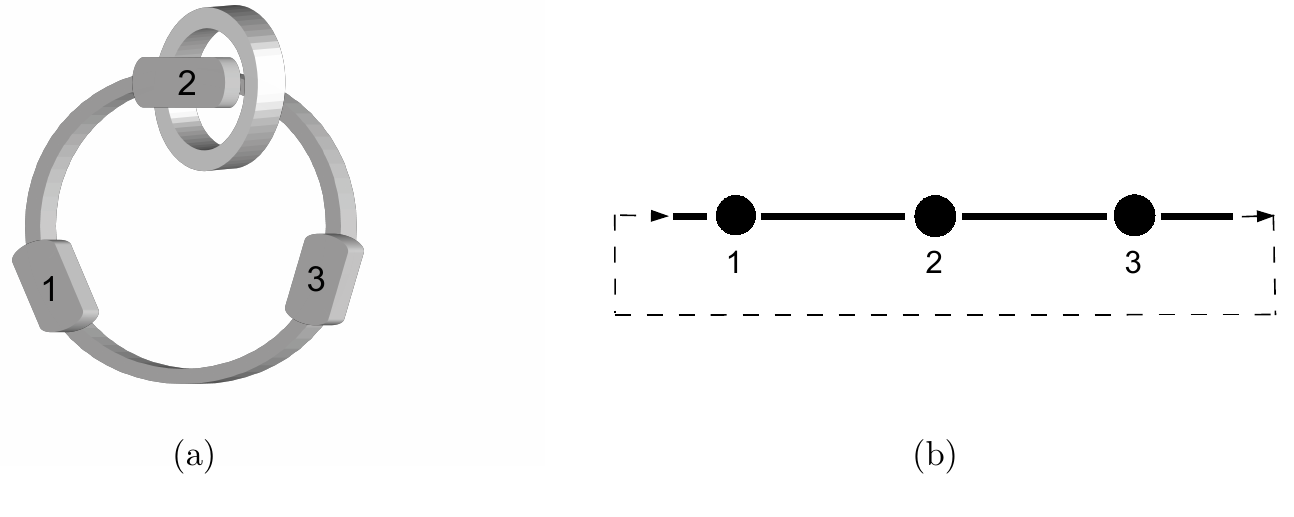}
	\caption{A [2]catenane and its representations. (a) Block diagram. (b) Hybrid model. Dashed line indicates periodic boundary condition.}
	\label{fig:catenane}
	\end{figure}

A {\it [2]catenane} is a supramolecular complex consisting of two mechanically interlocked rings, called macrocycles.
In the experiments of Ref.~\cite{Leigh2003} these macrocycles were unequal in size, and the smaller macrocycle had three binding sites, or {\it stations}, on the larger macrocycle.
In Fig.~\ref{fig:catenane}(a) these stations are represented schematically by numbered boxes.
This system has three metastable states, corresponding to the presence of the smaller macrocycle at station 1, 2 or 3.

Consider now the thermally activated motion of the small macrocycle along the large macrocycle.
In the discrete state model the state space of the system is depicted by three points, among which the macrocycle makes instantaneous transitions, representing sudden jumps from one station to another.
In our hybrid model, shown in Fig.~\ref{fig:catenane}(b), the discrete states are connected by continuous line segments, or {\it tracks}, along which the system performs diffusive motion during each transition.
The dashed line indicates periodic boundary conditions. 

\begin{figure}[tbp]
	\includegraphics[trim = 0in 0in 0in 0in, scale = 0.85]{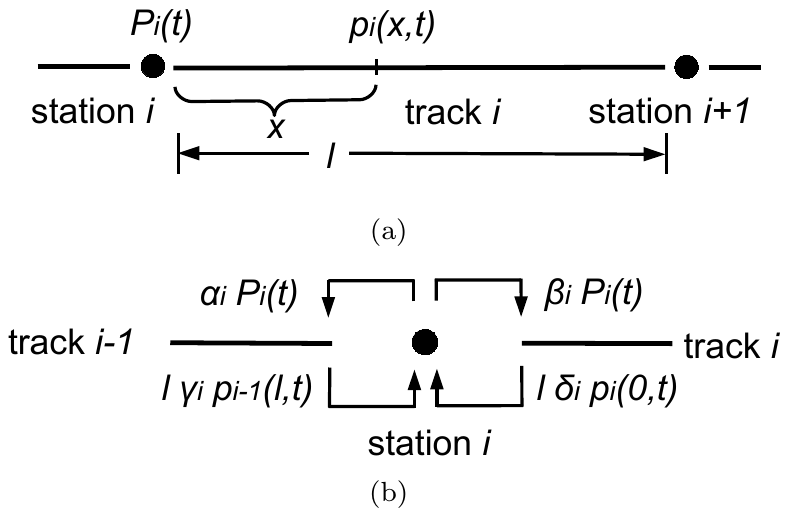}
	\caption{Elements of the hybrid model. (a) Probability at station $i$ is denoted by $P_i(t)$, and probability density on track $i$ by $p_i(x,t)$, where distance $x$ is measured from station $i$. Each track is assumed to be of length $l$. (b) Rate parameters of the model: $\alpha_i$ and $\beta_i$ denote the transition rates from station $i$ to tracks $(i-1)$ and $i$, respectively; $l \gamma_i$ and $l \delta_i$ are associated with the reverse transitions, as described in the text.}
	\label{fig:currents}
	\end{figure}

Let $P_i(t)$ denote the probability to find the system in station $i$ at time $t$, and let $p_i(x,t)$ be the probability density to find the system at a position $x$ along track $i$ at time $t$.
In our notation, a given track is designated by the same index as the station on its left; $x$ specifies the distance along a track; and for simplicity we assume each track to be of length $l$.
See Fig.~\ref{fig:currents}(a) for an illustration.
Because of the periodic nature of the state space we make the identifications: $i+1\equiv 1$ if $i = 3$ and $i-1 \equiv 3$ if $i = 1$.
The total probability is normalized to unity:
\begin{equation}
\sum_{i=1}^{n_S}  P_i(t) \,\, + \,\, \sum_{i=1}^{n_T} \int_0^l {\rm d}x \, p_i(x,t)  = 1 .
\end{equation}
Here, $n_S=3$ is the number of stations and $n_T=3$ is the number of tracks, but in general these need not be equal (see Sec.~\ref{sec:generalization}).

We now specify the dynamics of our model.
When the system is in station $i$, it has a probability per unit time $\alpha_i$ to make a leftward transition to the location $x=l$ on track $i-1$, and similarly a probability rate $\beta_i$ to make a rightward transition to the location $x=0$ on track $i$; see Fig.~\ref{fig:currents}(b).
When the system is on one of the tracks, it performs diffusive motion with a fixed diffusion constant $D$, with reflective (hard-wall) boundary conditions at $x=0$ and $x=l$.
Upon reaching either end of the track the system might jump into the adjacent station.
These jumps are characterized by probability rates $\gamma_i$ (for transitions from track $i-1$ to station $i$) and $\delta_i$ (from track $i$ to station $i$).
More precisely, the probability per unit time for the system to make a transition from track $i-1$ to station $i$ is given by the product $l\gamma_i p_{i-1}(l,t)$, where the factor $l$ is introduced on dimensional grounds, so that the parameter $\gamma_i$ has units of a probability rate i.e. $(\text{time})^{-1}$.
A similar expression holds for transitions from track $i$ to station $i$; see Fig.~\ref{fig:currents}(b).

These transitions give rise to a flow of probability between stations and adjacent tracks.
The net current from track $i-1$ to station $i$ is given by 
	\begin{equation}
	\label{eq:J>i}
	 J_{\rightarrow i}(t) = l \gamma_i p_{i-1}(l,t) - \alpha_i P_i(t),
	 \end{equation}
and that from station $i$ to track $i$ by
	\begin{equation}
	\label{eq:Ji>}
	J_{i \rightarrow}(t) =  \beta_i P_i(t) - l \delta_i p_i(0,t).
	\end{equation}
(The subscripts ``$\rightarrow i$'' and ``$i\rightarrow$'' indicate rightward probability current into and out of station $i$, respectively.)
The diffusive current at position $x$ along track $i$ is
	\begin{equation}
	\label{eq:Jd}
	J_i^d(x,t) = - D \frac{\partial}{\partial x} p_i(x,t),
	\end{equation}
and the reflective boundary conditions imply that $J_i^d(0,t) = J_i^d(l,t)=0$.
	
These currents generally lead to changes in the probability distribution. 
The rate of change of the probability to find the system at station $i$ is the difference between the incoming and the outgoing currents,
	\begin{equation}
	\label{eq:Pidot}
	\frac{\mathrm{d}P_i(t)}{\mathrm{d}t} = J_{\rightarrow i}(t) - J_{i \rightarrow}(t),
	\end{equation}
and that of probability density along track $i$ obeys a diffusion equation with a source and sink:
	\begin{eqnarray}
	\label{eq:pidot}
	\frac{\partial p_i(x,t)}{\partial t} & = & - \frac{\partial}{\partial x} J^d_i(x,t)  + \delta(x-0)J_{i \rightarrow}(t) \nonumber \\ 
	& & - \delta(x-l) J_{ \rightarrow i+1}(t).
	\end{eqnarray}
Eqs.~\ref{eq:Pidot} and \ref{eq:pidot} form a set of six coupled, linear equations (taking $i=1,2,3$) which collectively constitute the master equation describing the stochastic evolution of the system.

\section{Constraints Imposed by Detailed Balance}
	\label{sec:DB}

Since our model is meant to represent a system immersed in a thermal reservoir, the dynamics described by our master equation should have the property that when the rate parameters $\alpha_i$, $\beta_i$, $\gamma_i$ and $\delta_i$ are held fixed, the system relaxes to a state of equilibrium in which all currents are zero.
This condition of {\it detailed balance} imposes constraints on the rate parameters.
We now explore these constraints, and show that they allow us to rewrite the rate parameters in the suggestive form given by Eq.~\ref{eq:Arrhenius} below.

\begin{figure}[tbp]
	\includegraphics[trim = 0in 0in 0in 0in, scale = .85]{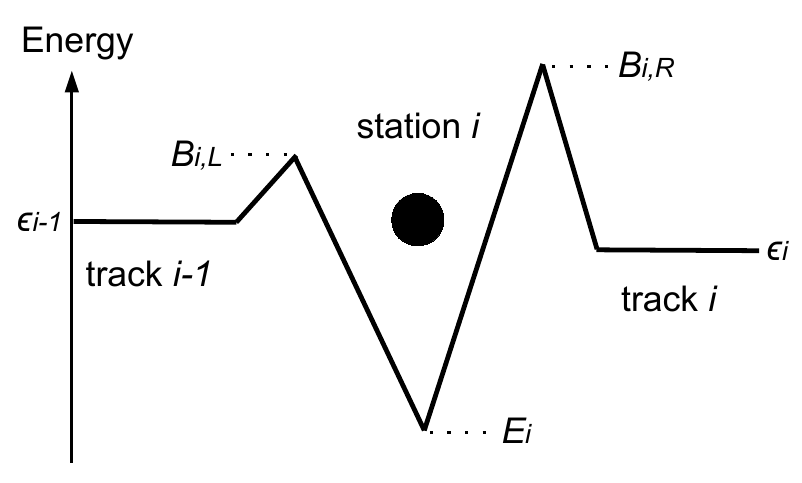}
	\caption{Free energies and barriers. Station $i$ has energy $E_i$, and track $i$ energy $\epsilon_i$. Track $(i-1)$ and station $i$ are separated by barrier $B_{i,L}$. Station $i$ and track $i$ by $B_{i,r}$.}
	\label{fig:energy}
	\end{figure}

Let $P_i^{eq}$ and $p_i^{eq}(x)$ denote, respectively, the station probabilities and track probability densities in the equilibrium state.
According to the condition of detailed balance, the currents appearing on the left sides of Eqs.~\ref{eq:J>i}, \ref{eq:Ji>} and \ref{eq:Jd} vanish when these values are substituted into their right sides.
This leads to the relations,
\begin{eqnarray}
	\label{eq:J>ieq}
	l \gamma_i p_{i-1}^{eq}(l) &=& \alpha_i P_i^{eq} \equiv \eta \exp[-B_{i,L}] \\
	\label{eq:Ji>eq}
	l \delta_i p_i^{eq}(0) &=& \beta_i P_i^{eq} \equiv \eta \exp[-B_{i,R}] \\
	\label{eq:Jdeq}
	\frac{\partial}{\partial x}p_i^{eq}(x) &=& 0.
\end{eqnarray}
Eqs.~\ref{eq:J>ieq} and \ref{eq:Ji>eq}, together with an arbitrary frequency scale $\eta$, define the dimensionless parameters $B_{i,L}$ and $B_{i,R}$, while Eq.~\ref{eq:Jdeq} implies that the equilibrium probability density is uniform along each track.
Introducing the dimensionless parameters
	\begin{equation}
	\label{eq:energy}
	E_i \equiv - \ln{P_i^{eq}} \quad {\rm and} \quad \epsilon_i \equiv - \ln{(l p_i^{eq})}
	\end{equation}
now allows us to rewrite Eqs.~\ref{eq:J>ieq} and \ref{eq:Ji>eq} as follows:
	\begin{equation}
	\label{eq:Arrhenius}
	\begin{split}
	\alpha_i =   \eta \, e^{-B_{i,L} + E_i} \quad, &  \quad \beta_i =   \eta \, e^{-B_{i,R}+ E_i}  ,\\
	\gamma_i =  \eta \, e^{-B_{i,L} + \epsilon_{i-1}} \quad, & \quad \delta_i =  \eta \,  e^{-B_{i,R}+\epsilon_i}.
	\end{split}
	\end{equation}
Because these expressions resemble the Arrhenius form for chemical reaction rates, it is natural to interpret $E_i$ (or $\epsilon_i$) as the energy of the small macrocycle when it is at station $i$ (or on track $i$); and $B_{i,L}$ (or $B_{i,R}$) as the height of the barrier that separates station $i$ from the track immediately to its left (or right).
These energies are given in units of $k T$, where $k$ is Boltzmann's constant and $T$ the absolute temperature of the environment.
We note that this interpretation, illustrated in Fig.~\ref{fig:energy}, is convenient but not strictly necessary: the NPT that we derive below is an exact result stated in terms of the well-defined quantities $E_i$, $\epsilon_i$, $B_{i,L}$ and $B_{i,R}$, regardless of whether or not we choose to interpret these as energies and barriers.

\section{Statement and Proof of NPT}
	\label{sec:NPT}
	
Stochastic pumping corresponds to the periodic variation of the transition rates $\{\alpha_i,\beta_i,\gamma_i,\delta_i\}$, subject to the constraints imposed by detailed balance.  In the energetic picture introduced above, this translates to the periodic variation of the energies and barriers $\{E_i,\epsilon_i,B_{i,L},B_{i,R}\}$, that is, $E_i(t+\tau) = E_i(t)$, etc., where $\tau$ is the period of the pumping.
Under these conditions the system relaxes to a unique periodic steady state,
	\begin{equation}
	\label{eq:periodic}
	P_i^{ps}(t + \tau) = P_i^{ps}(t) \quad , \quad p_i^{ps}(x,t+\tau) = p_i^{ps}(x,t),
	\end{equation}
characterized by time-periodic currents passing through the stations, $J_{\rightarrow i}^{ps}(t)$ and $J_{i \rightarrow}^{ps}(t)$, and along the tracks, $J^{d, ps}_i (x,t)$.  We are interested in the {\it integrated} currents,
	\begin{eqnarray}
	\label{eq:Phi}
	\Phi_{\rightarrow i}^{ps} \equiv \int_\tau \mathrm{d}t \, J_{\rightarrow i}^{ps}(t), \quad
	\Phi_{i \rightarrow}^{ps} \equiv \int_\tau \mathrm{d}t \, J_{i \rightarrow}^{ps}(t), \nonumber \\
	\Phi^{d, ps}_i(x) \equiv \int_{\tau} \mathrm{d}t \, J^{d, ps}_i (x,t), \quad \quad
	\end{eqnarray}
where $\int_\tau$ denotes an integral over one period of pumping.  Here, $\Phi_{\rightarrow i}^{ps}$ represents the net flow of probability from track $i-1$ into station $i$  over one pumping cycle, and $\Phi_{i \rightarrow}^{ps}$ and $\Phi^{d, ps}_i(x)$ have similar interpretations; the integrated currents thus measure the extent to which the pumping of the energies and barriers drives a non-zero current around the cycle depicted in Fig.~\ref{fig:catenane}(b).
Physically, these currents measure our ability to generate directed mechanical motion of the small macrocycle around the large macrocycle, by the periodic variation of external parameters, with positive and negative $\Phi$'s corresponding to clockwise and counterclockwise motion, respectively; see Fig.~\ref{fig:catenane}(a).

These considerations apply to the time-periodic pumping of any combination of the parameters $\{E_i,\epsilon_i,B_{i,L},B_{i,R}\}$.
In the subsequent analysis, however, we will assume that the track energies $\epsilon_i$ remain constant with time, while the station and/or barrier energies (the $E$'s and $B$'s) are varied periodically.
Thus we treat the tracks as fixed conduits for diffusive motion from one station to another;
this is in keeping with the relevant experimental studies \cite{Leigh2003, Panman2010}, which did not include any time-dependent track energies.
The no-pumping theorem (NPT) that we now prove states that, in order to generate non-zero integrated currents, we must vary some combination that includes {\it both} station energies (the $E_i$'s) {\it and} barrier energies (the $B_{i,L}$'s and/or $B_{i,R}$'s).
In other words: (1) if we vary only the barrier energies, while keeping the station energies fixed, or (2) if we vary only the station energies, while keeping the barrier energies fixed, then in either case all the integrated currents will be zero.
To prove the NPT we now consider these cases separately.

The first case is easy to analyze, and intuitively plausible.
Let the term {\it instantaneous equilibrium distribution} denote the equilibrium distribution corresponding to the current parameter values, at any instant in time.
By  Eq.~\ref{eq:energy}, this distribution depends only on the state energies and not on the barrier energies.
Thus when the $E_i$'s and $\epsilon_i$'s are held fixed, the instantaneous equilibrium distribution $\{P_i^{eq},p_i^{eq}(x)\}$ also remains fixed.
Moreover, by Eqs.~\ref{eq:Pidot} and \ref{eq:pidot} this time-independent distribution is a stationary solution of the dynamics, even if the barrier energies are varied  with time.
At the same time, any time-periodic pumping protocol uniquely determines a time-periodic solution of the dynamics.
It follows that when the state energies are held fixed and the barrier energies are varied periodically, then the unique periodic steady state of the system is simply the time-independent instantaneous equilibrium distribution.
In this state, all the instantaneous currents are zero, and therefore the integrated currents also vanish.

	
Now consider the situation in which the station energies are varied periodically in time, $E_i(t+\tau) = E_i(t)$, and the barriers $\{B_{i,L}, B_{i,R}\}$ (together with the track energies $\epsilon_i$) are kept fixed.
The NPT then follows from a combination of two conditions: the \emph{detailed balance} constraints, Eq. \ref{eq:Arrhenius}, and the \emph{periodicity} of the probability distribution, Eq. \ref{eq:periodic}, as we now show.

Let us first explore the consequences of the detailed balance constraints.
Combining Eq.~\ref{eq:Arrhenius} with the expressions for the instantaneous currents, Eqs. \ref{eq:J>i} and \ref{eq:Ji>}, we derive for each station $i$,
	\begin{equation}
	\label{eq:Jeq}
	e^{B_{i,L}} J_{\rightarrow i}(t) + e^{B_{i,R}} J_{i \rightarrow}(t) = \eta \,  l   [e^{\epsilon_{i-1}} p_{i-1}(l, t) - e^{\epsilon_i} p_i(0,t)  ].
	\end{equation}
Note that the superscript {\it ps} does not appear here, as this relation is valid whether or not the system has reached a periodic steady state.
Summing both sides over $i$, we get
	\begin{equation}
	\label{eq:Jeqsum}
	\sum_{i} [ e^{B_{i,L}} J_{ \rightarrow i}(t) + e^{B_{i,R}} J_{i \rightarrow}(t)] = \eta \, l \sum_i e^{\epsilon_i} [  p_i(l, t) - p_i(0,t) ].
	\end{equation}
The definition of the diffusive current, Eq. \ref{eq:Jd}, implies
	\begin{equation}
	\label{eq:Jdint}
	-\frac{1}{D} \int_0^l \mathrm{d}x \, J_i^{d}(x,t) = \Big[ p_i(l,t) - p_i(0,t)\Big],
	\end{equation}
which combines with Eq. \ref{eq:Jeqsum} to give
	\begin{equation}
	\label{eq:Jeqsumfinal}
	\sum_{i} \bigg[ e^{B_{i,L}} J_{\rightarrow i}(t) + e^{B_{i,R}} J_{i \rightarrow}(t)  + \frac{1}{D} e^{\epsilon_i} \int_0^l \mathrm{d}x \, J_i^{d}(x,t) \bigg] = 0.
	\end{equation}
(See Refs. \cite{Chernyak2009JCP,Mandal2011} for the analogue of this relation in the context of a discrete state model.)
If we now assume the system has reached a periodic steady state, and we integrate this relation over one period, we get
	\begin{equation}
	\label{eq:Phieq}
	\sum_{i} \bigg[ e^{B_{i,L}} \Phi_{\rightarrow i}^{ps} + e^{B_{i,R}} \Phi_{i \rightarrow}^{ps}  + \frac{\eta \, l}{D} e^{\epsilon_i} \int_0^l \mathrm{d}x \, \Phi_i^{d,ps}(x) \bigg] = 0.
	\end{equation}

Now we explore the implications of the periodicity of the probability distribution, Eq. \ref{eq:periodic}.
Since the probability to find the system in station $i$ returns to the same value after each period, the integrated current that enters that station from the left must be balanced by the integrated current that exits from the right: $\Phi_{\rightarrow i}^{ps} = \Phi^{ps}_{i \rightarrow}$.
This value is in turn equal to the integrated current entering track $i$ from the left.
Along track $i$ the integrated current $\Phi^{d,ps}_i(x)$ must be the same for any two points $x_1$ and $x_2$, otherwise there would be a net accumulation or depletion of probability in the interval between those points, over each period.
Proceeding in this manner around the entire circuit we conclude that the integrated current is uniform all along:\footnote{A derivation of Eq.~\ref{eq:Phiperiodic} directly from the master equations requires a more careful treatment, in which the source and sink terms in Eq. \ref{eq:pidot} are displaced slightly from the track-ends. As this does not contribute conceptually to the main line of the proof we present the analysis in Appendix~\ref{app:epsilon}.}
	\begin{equation}
	\label{eq:Phiperiodic}
	\Phi_{\rightarrow i}^{ps} = \Phi^{ps}_{i \rightarrow} = \Phi^{d,ps}_i(x) = \Phi_{\rightarrow i+1}^{ps} \cdots \equiv \Phi.
	\end{equation}
It follows immediately from Eqs.  \ref{eq:Phieq} and \ref{eq:Phiperiodic} that all integrated currents are zero.
Thus the NPT is established for the three-state model depicted in Fig.~\ref{fig:catenane}.

\section{Generalizations}
	\label{sec:generalization}

We now generalize our discussion along two different directions.
First, following Ref.~\cite{Panman2010}, we allow for a spatially non-uniform (but still time-independent) free energy landscape along each of the tracks.
Secondly, we move beyond the simple three-station, three-track network shown in Fig.~\ref{fig:catenane}, and extend our model to encompass an arbitrary, finite network of stations and tracks.
The analysis involved in these more general situations is similar to that presented in Sec.~\ref{sec:NPT}, therefore to avoid repetition we sketch only the key ideas in the following discussion.
	
First we allow a nonuniform energy landscape $V_i(x)$ along each track $i$, instead of constant $\epsilon_i$, again in units of $k T$; this leads to the expressions, 
	\begin{eqnarray}
	\label{eq:Jdmod}
	J_i^d(x,t)  =  - D \bigg[ \frac{\partial p_i(x, t)}{\partial x} + \frac{\partial V_i(x)}{\partial x} p_i(x,t)  \bigg] \\
	\gamma_i  =  \eta \, e^{- B_{i,L} + V_{i-1}(l)} \quad , \quad \delta_i  =  \eta \, e^{- B_{i,R} + V_i(0)}.
	\end{eqnarray}
(compare Eqs.~\ref{eq:Jd} and \ref{eq:Arrhenius}, respectively).
When all the state energies are held fixed and only the barrier energies are varied with time, the arguments presented earlier apply here without modification, and we can  conclude that all currents vanish in the steady state.
When instead the barrier energies are fixed and the station energies are varied periodically, in place of Eq. (\ref{eq:Jeq}) we have
	\begin{eqnarray}
	\label{eq:JeqV}
	e^{B_{i,L}} J_{ \rightarrow i}^{ps}(t) + e^{B_{i,R}} J_{i \rightarrow}^{ps}(t) \qquad \qquad \qquad \nonumber \\ 
	 = \eta \, l \bigg[ e^{V_{i-1} (l)} p_{i-1}^{ps}(l, t) - e^{V_i(0)}p_i^{ps}(0,t) \bigg],
	\end{eqnarray}
which leads to a modified version of Eq. (\ref{eq:Phieq}),
	\begin{equation}
	\label{eq:PhieqsumfinalV}
	\sum_{i} \bigg[ e^{B_{i,L}} \Phi_{ \rightarrow i}^{ps} + e^{B_{i,R}} \Phi_{i \rightarrow }^{ps}  + \frac{\eta \, l}{D} \int_0^l \mathrm{d}x \, e^{V_i(x)} \Phi_i^{d,ps}(x) \bigg] = 0.
	\end{equation}
The periodicity of the probability distribution, Eq.~\ref{eq:periodic}, again implies a uniform integrated current, Eq.~\ref{eq:Phiperiodic}.
The combination of Eqs.~\ref{eq:Phiperiodic} and \ref{eq:PhieqsumfinalV} in turn immediately implies that all integrated currents vanish, and thus the NPT is established.

	\begin{figure}[tbp]
	\includegraphics[trim = 0in 0in 0in 0in, scale = 0.85]{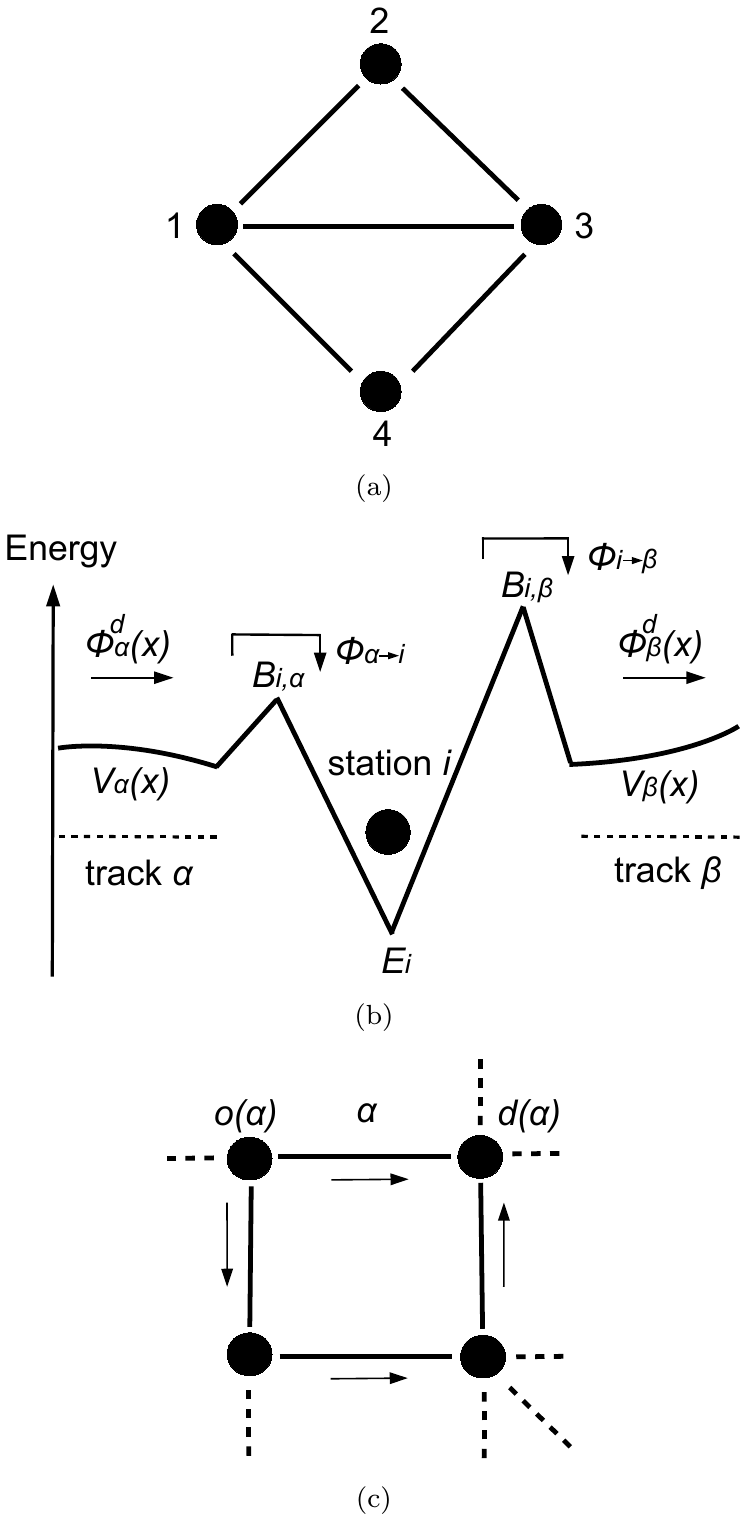}
	\caption{Aspects of generic networks. (a) A network with $n_S = 4$ stations and $n_T = 5$ tracks.  (b) Notations for a generic network, as described in the text.  Although we show a station connected to two tracks, in general a station can be adjacent to as few as one and as many as $n_T$ tracks. (c) A single cycle within a network.  Arrows indicate the (arbitrarily assigned) directions of the tracks, and the origin and destination of track $\alpha$ are explicitly labelled.}
	\label{fig:currentsgen}
	\end{figure}
	
Now consider a more general network, composed of $n_S$ stations and $n_T$ tracks,, which need not be equal.
We assume this network is {\it connected}: for any pair of stations $i$, $j$, there exists at least one path -- a sequence of alternating tracks and stations -- that connects station $i$ to station $j$.
(This assumption is not restrictive, as any finite network can be decomposed into two or more connected networks, each of which satisfies the NPT.)
Fig.~\ref{fig:currentsgen}(a) illustrates a connected network, with $n_S=4$ and $n_T=5$.
The [2]catenane model analyzed in Sections \ref{sec:model} - \ref{sec:NPT} had only one \emph{cycle} -- a closed loop of stations and tracks (see Fig.~\ref{fig:catenane}) -- but in the more general case considered here there can be more than one cycle, as shown in Fig.~\ref{fig:currentsgen}(a).
	
We will now use Roman indices $i \in\{1, \ldots n_S \}$ to denote stations and Greek indices $\alpha \in \{ 1, \ldots n_T \}$ to denote tracks.
To each track $\alpha$ we assign an arbitrary direction, such that the station at one end of the track is viewed as the \emph{origin} $o(\alpha)$ and the station at the other end is the \emph{destination} $d(\alpha)$, as illustrated in Fig. \ref{fig:currentsgen}(c).
These assignments are made independently for the various tracks, and a given station can simultaneously serve as the origin of some tracks and the destination of others.
Station and track energies are denoted by $E_i$ and $V_\alpha(x)$, respectively; the barrier between station $i$ and track $\alpha$ is given by $B_{i, \alpha}$; the integrated current from track $\alpha$ (or station $i$) to station $i$ (or track $\alpha$) is given by $\Phi^{ps}_{\alpha \rightarrow i}$ (or $\Phi^{ps}_{i \rightarrow \alpha}$); and the integrated diffusive current on track $\alpha$ along the preassigned direction by $\Phi^{d,ps}_\alpha(x)$.
These conventions are illustrated in Fig.~\ref{fig:currentsgen}(b), where superscripts ``ps" have been omitted to avoid clutter.

As earlier, under the time-periodic variation of station and/or barrier energies the system settles into a unique periodic steady state.
Moreover, when all the state energies are held fixed and only the barrier energies are varied with time, then the entire system simply relaxes to a state of thermal equilibrium, in which all currents vanish.
To establish the NPT, it remains to show that the integrated current vanishes when the barrier energies are kept fixed and the station energies are varied periodically.

The periodicity of the probability distribution implies, for each station $i$,
	\begin{equation}
	\label{eq:periodicgeni}
	\sum_{\{\alpha | d(\alpha) = i\}} \Phi^{d,ps}_{\alpha \rightarrow i} =
	\sum_{\{\alpha | o(\alpha) = i\}} \Phi^{d,ps}_{i \rightarrow \alpha};
	\end{equation}
the sum on the left represents the net integrated current into station $i$ from all tracks $\alpha$ for which it is the destination, and the sum on the right is the net integrated current out of the station into all tracks to which it is the origin.
For each track $\alpha$ we have
	\begin{equation}
	\label{eq:periodicgenalpha}
	\Phi_{o(\alpha) \rightarrow \alpha}^{ps} = \Phi_\alpha^{d, ps}(x) = \Phi_{\alpha \rightarrow d(\alpha)}^{ps}
	\end{equation}
(compare with Eq.~\ref{eq:Phiperiodic}),
which ensures that probability neither accumulates nor depletes anywhere on the track, with each cycle.
Eqs.~\ref{eq:periodicgeni} and \ref{eq:periodicgenalpha} have an interesting consequence: if there exists a non-zero integrated current in the system, then it must be part of a cycle (of alternating stations and currents) along which all of the integrated currents point in the same direction.
The intuition is straightforward: to prevent the systematic accumulation of probability within the network, current must flow in a circle.
We now formalize and establish this statement, and then use it to prove that all integrated currents must be zero (when the barriers are fixed and the station energies are varied periodically), following arguments similar to those presented in Ref.~\cite{Mandal2011}.

Without loss of generality, suppose that a particular track $\alpha$ supports a positive integrated current: $\Phi_\alpha^{d,ps} > 0$.
(Equivalent arguments would apply if the sign were negative.)
Eq.~\ref{eq:periodicgenalpha} then implies positive integrated currents $\Phi_{o(\alpha) \rightarrow \alpha}^{ps}$ and $ \Phi_{\alpha \rightarrow d(\alpha)}^{ps}$. 
As probability cannot deplete on station $o(\alpha)$ over a complete cycle, $o(\alpha)$ must have neighboring track(s) $\beta$ such that $\Phi^{ps}_{\beta \rightarrow o(\alpha)} > 0$.
Similarly, to avoid accumulation of probability, $d(\alpha)$ must have neighboring track(s) $\gamma$ such that $\Phi_{d(\alpha) \rightarrow \gamma} > 0$.
The periodic conservation of probability in turn establishes the directionality of the integrated currents along these tracks: on track $\beta$, the integrated current must flow toward station $o(\alpha)$, and on track $\gamma$ it must flow away from station $d(\alpha)$. 
Continuing in this manner, we now construct a set $D(\alpha)$ of stations and tracks to which there is a positive flow of current from $\alpha$; this will consist of $d(\alpha), \gamma,$ so on. 
Similarly, we construct set $S(\alpha)$ of stations and tracks from which there is a positive flow to $\alpha$; this will consist of $o(\alpha), \beta,$ so on.
In order to prevent the accumulation of probability in the former set and its depletion in the latter with each complete cycle,
$D(\alpha)$ and $S(\alpha)$ must have a common element.
This implies the existence of a cycle 
	\begin{equation}
	\label{eq:cycle}
	c \equiv \alpha \rightarrow d(\alpha) \rightarrow \gamma \rightarrow \ldots \rightarrow \beta \rightarrow o(\alpha) \rightarrow \alpha,
	\end{equation}
along which all integrated currents flow in the same direction.

Now recall that each track in our network has been assigned a direction, pointing from its origin to its destination.
By assumption, for track $\alpha$ this direction is parallel to the direction of probability flow around the cycle $c$, indicated by Eq.~\ref{eq:cycle}.
However, since the assignment of track directions is arbitrary, each of the remaining tracks in the cycle ($\beta, \cdots \gamma$) might be directed either parallel or anti-parallel to the flow along the cycle.
Let us therefore introduce a factor $s_\mu=\pm 1$, defined for every track $\mu$ in the cycle $c$, such that $s_\mu=+1$ (or $-1$) if track $\mu$ is oriented parallel (or anti-parallel) to the flow in the cycle.
Then $\Psi^{d,ps}_\mu \equiv s_\mu \Phi^{d,ps}_\mu>0$ for each track $\mu$ in the cycle $c$, and $\Psi^{ps}_{o(\mu) \rightarrow \mu} \equiv s_\mu \Phi^{ps}_{o(\mu) \rightarrow \mu}$ and $\Psi^{ps}_{\mu \rightarrow d(\mu)} \equiv s_\mu \Phi^{ps}_{\mu \rightarrow d(\mu)}$ are positive as well, by Eq.~\ref{eq:periodicgenalpha}. 
The analogue of Eqs. \ref{eq:Phieq} and \ref{eq:PhieqsumfinalV} for the cycle $c$ is given by
	\begin{eqnarray}
	\label{eq:PsieqsumfinalVgen}
	\sum_{\mu} \bigg[ e^{B_{o(\mu), \mu}} \Psi_{ o(\mu) \rightarrow \mu}^{ps}  + e^{B_{d(\mu), \mu}} \Psi_{ \mu \rightarrow d(\mu)}^{ps} \nonumber \\ + \frac{\eta \, l}{D} \int_0^l \mathrm{d}x \, e^{V_\mu(x)} \Psi_\mu^{d,ps}(x)  \bigg] = 0,
	\end{eqnarray}
where the sum is taken over all tracks $\mu$ in cycle $c$.
Since all the $\Psi$'s are positive, Eq.~\ref{eq:PsieqsumfinalVgen} cannot be satisfied.
Hence our starting assumption, the existence of a positive integrated current along track $\alpha$, must be invalid.
This establishes the no-pumping theorem.

\section{Conclusion}
	\label{sec:conclusion}

Motivated by the experiments of Refs.~\cite{Panman2010,Panman2012} we have introduced a hybrid class of models for artificial molecular machines, combining elements of both discrete and continuous models.
Within this picture, we have established the validity of the no-pumping theorem~\cite{Rahav2008, Horowitz2009}, which had earlier been derived for entirely discrete and entirely diffusive models.
We illustrated our derivation in detail for the [2]-catenane complex studied in Ref~\cite{Leigh2003}, and then generalized it to more complicated situations.
It remains to be investigated whether other results~\cite{Chernyak2008, Chernyak2009JCP, Sinitsyn2009, Ren2011, Sinitsyn2011} that have been derived for purely discrete or purely continuous models, apply to our hybrid model.

	   \acknowledgments
	   We thank Andy Ballard, Shaon Chakrabarti, Sebastian Deffner, Zhiyue Lu and Haitao Quan for useful discussions, and gratefully acknowledge financial support from the National Science Foundation (USA) under grants 0925365 and 0906601, the U.S.-Israel Binational Science Foundation under grant 2010363, and the University of Maryland, College Park.

	\appendix
	
	\section{Derivation of Eq. \ref{eq:Phiperiodic}}
	\label{app:epsilon}
	
	\begin{figure}[tbp]
	\includegraphics[scale = 0.85]{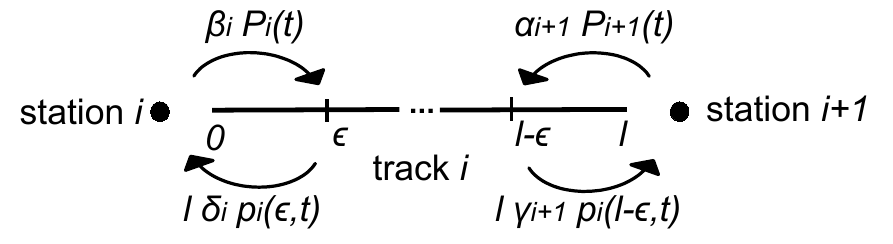}
	\caption{Source and sink terms are moved away from the track-ends by a parameter $0 < \epsilon <l/2$.}
	\label{fig:epsilon}
	\end{figure}
	
Here we establish Eq.~\ref{eq:Phiperiodic} directly from the master Eqs.~\ref{eq:Pidot} and \ref{eq:pidot}, evaluated in the periodic steady state, Eq.~\ref{eq:periodic}.
We begin by assuming that the source and sink terms in Eq.~\ref{eq:pidot} are displaced away from the track-ends, as illustrated in Fig.~\ref{fig:epsilon}; Eq.~\ref{eq:pidot} then becomes
	\begin{equation}
	\label{eq:pidotapp}
	\frac{\partial p_i(x,t)}{\partial t} = - \frac{\partial}{\partial x} J^d_i(x,t)  + \delta(x-\epsilon)J_{i \rightarrow}(t) - \delta(x-l+\epsilon) J_{ \rightarrow i+1}(t),
	\end{equation}
where $\epsilon > 0$.
We recover the situation described in the main body of the text in the limit $\epsilon\rightarrow 0$.

In the periodic steady state we have $P_i^{ps}(t + \tau) = P_i^{ps}(t)$ for each station $i$.
Equivalently, $\int_\tau \mathrm{d}t \, \mathrm{d} P_i^{ps}(t) / \mathrm{d} t = 0$, which from Eqs.~\ref{eq:Pidot} and \ref{eq:Phi} implies
	\begin{equation}
	\label{eq:periodiciapp}
	\Phi^{ps}_{\rightarrow i} = \Phi^{ps}_{i \rightarrow}.
	\end{equation}
This is the first part of Eq.~\ref{eq:Phiperiodic}.
Similarly, for each track $i$ we have $\int_\tau \mathrm{d}t \, \partial p_i^{ps}(x,t) / \partial t = 0$, which combines with Eq.~\ref{eq:pidotapp} and \ref{eq:Phi} to give
	\begin{equation}
	\label{eq:Phiprime}
	\frac{\partial}{\partial x} \Phi_i^{d, ps}(x) = \delta(x - \epsilon) \Phi_{i \rightarrow }^{ps} - \delta(x - l + \epsilon) \Phi_{\rightarrow i+1}^{ps}.
	\end{equation}
From the reflective boundary conditions,  $J_i^d(0,t) = J_i^d(l,t)=0$, we have
	\begin{equation}
	\label{eq:Phiprimebdy}
	\Phi_i^{d, ps} (0) = \Phi_i^{d, ps}(l) = 0.
	\end{equation}
Finally, solving Eq.~\ref{eq:Phiprime} with boundary conditions Eq.~\ref{eq:Phiprimebdy} we get 
	\begin{equation}
	\label{eq:pss_p_sol}
	\Phi^{d,ps}_i (x) = \left \{
		\begin{array}{ll}
		0 & \quad \text{for} \quad 0 \leq x < \epsilon,\\
		\Phi^{ps}_{i \rightarrow} = \Phi^{ps}_{\rightarrow i+1} & \quad \text{for} \quad \epsilon < x < l - \epsilon,\\
		0 & \quad \text{for} \quad l - \epsilon < x \leq l.
		\end{array}
	\right.
	\end{equation}
Eqs.~\ref{eq:periodiciapp} and \ref{eq:pss_p_sol}, together with the limit $\epsilon\rightarrow 0$, lead to Eq.~\ref{eq:Phiperiodic}.

	\end{document}